\documentclass[10pt,conference]{IEEEtran}
\IEEEoverridecommandlockouts

\usepackage{enumitem}
\usepackage[ruled,vlined]{algorithm2e}
\usepackage{pgfplots}
\pgfplotsset{compat=1.18}

\usepackage{booktabs}
\usepackage[table,xcdraw]{xcolor}
\usepackage{siunitx}
\usepackage{float}
\usepackage{flushend}
\usepackage{tcolorbox}
\usepackage{lipsum} 
\usepackage{booktabs} 
\usepackage{multirow} 
\usepackage{array}    
\usepackage{graphicx}
\usepackage{xcolor}

\usepackage[table]{xcolor}  
\usepackage{cite}
\usepackage{amsmath,amssymb,amsfonts}
\usepackage{algorithmic}
\usepackage{textcomp}

\setlist[itemize]{leftmargin=*}
\setlist[enumerate]{leftmargin=*}
\newlist{steps}{enumerate}{1}
\setlist[steps, 1]{label = \textbf{RQ\arabic*.}}

\definecolor{mediumdarkblue}{rgb}{0,0,0.6} 
\definecolor{warmdarkred}{rgb}{0.6,0.05,0.05} 
\definecolor{darkred}{rgb}{0.5,0,0} 
\definecolor{darkblue}{rgb}{0,0,0.5} 
\definecolor{darkgreen}{rgb}{0,0.5,0}

\usepackage[colorlinks=true,
            linkcolor=mediumdarkblue,    
            citecolor=darkgreen,   
            filecolor=magenta, 
            urlcolor=cyan]{hyperref} 

\def\BibTeX{{\rm B\kern-.05em{\sc i\kern-.025em b}\kern-.08em
    T\kern-.1667em\lower.7ex\hbox{E}\kern-.125emX}}
\begin{document}

\title{\LARGE{\textit{Bridging Natural Language and Formal Specification}--} \\ \huge{Automated Translation of Software Requirements to LTL \\ via Hierarchical Semantics Decomposition Using LLMs}}

\author{
    \IEEEauthorblockN{
        Zhi Ma\textsuperscript{1},
        Cheng Wen\textsuperscript{2},
        Zhexin Su\textsuperscript{1},
        Xiao Liang\textsuperscript{1},
        Cong Tian\textsuperscript{1*} \thanks{*Corresponding author: Cong Tian.},
        Shengchao Qin\textsuperscript{1} and
        Mengfei Yang\textsuperscript{3}
    }
    \IEEEauthorblockA{
        \textsuperscript{1}School of Computer Science and Technology, Xidian University, Xi'an, China \\
        \textsuperscript{2}Guangzhou Institute of Technology of Xidian University, Guangzhou, China \\
        \textsuperscript{3}China Academy of Space Technology, Beijing, China \\
        \{mazhi, wencheng, qinshengchao\}@xidian.edu.cn, \{23031212487, xiaoliang\}@stu.xidian.edu.cn, \\ 
        ctian@mail.xidian.edu.cn, yangmf@bice.org.cn
    }
}

\maketitle

\begin{abstract}
Automating the translation of natural language (NL) software requirements into formal specifications remains a critical challenge in scaling formal verification practices to industrial settings, particularly in safety-critical domains. 
Existing approaches, both rule-based and learning-based, face significant limitations.
While large language models (LLMs) like GPT-4o demonstrate proficiency in semantic extraction, they still encounter difficulties in addressing the complexity, ambiguity, and logical depth of real-world industrial requirements.
In this paper, we propose \textsc{Req2LTL}, a modular framework that bridges NL and Linear Temporal Logic (LTL) through a hierarchical intermediate representation called \textit{OnionL}. 
\textsc{Req2LTL} leverages LLMs for semantic decomposition and combines them with deterministic rule-based synthesis to ensure both syntactic validity and semantic fidelity. 
Our comprehensive evaluation demonstrates that \textsc{Req2LTL} achieves 88.4\% semantic accuracy and 100\% syntactic correctness on real-world aerospace requirements, significantly outperforming existing methods. 
\end{abstract}


\begin{table*}[h]
  \vspace{-8pt}
  \centering
  \setlength{\abovecaptionskip}{0pt}
  \setlength{\belowcaptionskip}{0pt}
  \caption{Natural Language and Corresponding LTL Specifications} 
  \label{tab:nl_examples_ltl} 
    \resizebox{0.7\textwidth}{!}{%
    \begin{tabular}{ll}
      \toprule
      \textbf{Natural Language} & \textbf{LTL Formula} \\
      \midrule
      Once red, the light cannot become green next. & $G (\textnormal{red} \rightarrow X \neg \textnormal{green})$ \\[1pt]
      Once the light is red, it must remain red until it turns yellow. & $G (\textnormal{red} \rightarrow \textnormal{red} \, U \, \textnormal{yellow})$ \\[1pt]
      If b holds, next c holds until a holds or always c holds. & $G(\textnormal{b} \rightarrow X ((\textnormal{c} \, U \, \textnormal{a}) \lor G \textnormal{c}))$ \\[1pt]
      If a holds then c is true until b. & $G(\textnormal{a} \rightarrow (\textnormal{c} \, U \, \textnormal{b}))$ \\[1pt]
      Navigate to the green room while avoiding landmark 1. & $(F \textnormal{ green}) \land G (\neg \textnormal{ landmark 1})$ \\[3pt]
      Swing by landmark 1 before ending up in the red room. & $F (\textnormal{landmark 1} \land F \textnormal{ red})$ \\
      \bottomrule
    \end{tabular}%
  } 
\end{table*}

\section{Introduction}
The accurate and automated translation of natural language (NL) software requirements into formal language (FL) program specifications is crucial for leveraging formal verification techniques in industrial applications, particularly within safety-critical domains such as aerospace~\cite{paul2023formal}, operating systems~\cite{SPARC,CertiKOSAEGu2016}, compilers~\cite{leroy2016compcert}, and embedded controllers~\cite{Backes2019ReachabilityAF}.
Linear Temporal Logic (LTL) is widely used to formally specify temporal behaviors of reactive and embedded systems due to its precision in expressing complex safety and liveness properties and the availability of powerful verification tools (\textit{e.g.}, NuSMV~\cite{Cimatti2002NuSMV2A}, Spot~\cite{DuretLutz2016Spot2}). 
However, the current industry practice predominantly relies on human experts with deep domain and formal reasoning knowledge to manually translate natural language requirements into LTL formulas. 
This process is both time-consuming and prone to errors~\cite{Phipathananunth2022UsingMT,Liu2023EnhancingTC,Mashkoor2021ValidationOA,wen2024enchanting}.

Existing approaches aimed at automating NL-to-LTL translation are either rule-based or learning-based, each exhibiting significant drawbacks. 
Rule-based methods~\cite{Blasi2018TranslatingCCref3,Tan2012tCommentT_ref48javaJ} generally lack flexibility and are limited by variability and narrow scope.
In contrast, learning-based approaches~\cite{Pan2023DataEfficientLO_ref29,Rongali2022TrainingNSref32,Chen2023NL2TLTN} require extensive labeled datasets and often struggle to generalize beyond their training examples.
Recent advancements in large language models (LLMs), such as GPT-4o~\cite{openai2024gpt4o}, have shown potential in related domains like code generation and logical inference~\cite{Nijkamp2022CodeGenAOref33,wen2024automatically,cao2025informal,wen2024enchanting,du2024evaluating,su2024cfstra}, yet directly applying these models to complex NL-to-LTL translation tasks remains problematic due to the implicit temporal semantics, deeply nested logical structures, and context-specific constraints inherent in industrial requirements.

Three primary challenges limit the effectiveness of applying LLMs directly to this translation task. 
First, natural language often conveys temporal semantics implicitly through nuanced expressions (\textit{e.g.}, \textit{will be set} or \textit{unless}), which require deeper semantic interpretation beyond surface-level syntactic analysis. 
Second, industrial requirements often involve deeply nested logical constructs, including multiple conditionals, exceptions, and working mode-based constraints within single statements, posing significant challenges to the attention allocation and structural fidelity of LLMs.
Third, LLMs are autoregressive and prone to error compounding: if a partial formula is generated incorrectly, subsequent completions are likely to deviate further from the intended semantics. These challenges are exacerbated by the absence of structural validation mechanisms in most LLM-based methods, resulting in silent failures that are difficult to detect or rectify.

To effectively address these issues, we propose a novel intermediate representation called \textit{OnionL}, specifically designed to bridge the gap between natural language semantics and formal logic. 
\textit{OnionL} is a tree-structured intermediate language that reflects the compositional syntax of LTL while maintaining semantic alignment with domain-specific expressions. It provides a hierarchical, structured representation of requirements by explicitly modeling \textit{scopes}, \textit{relations}, and \textit{atomic propositions}. 
This structured abstraction serves as a semantic bridge, allowing LLMs to focus on local semantic extraction while delegating global logical composition to rule-based synthesis.

Building on the \textit{OnionL} representation, we propose the \textsc{Req2LTL} framework, which comprises two primary components: (1) the \textsc{Req2OnionL} module utilizes prompt-guided LLMs to decompose natural language requirements into structured \textit{OnionL} trees hierarchically; (2) the \textsc{OnionL2LTL} module employs a deterministic, rule-based approach to validate and translate these \textit{OnionL} structures into correct and standardized LTL formulas.
The entire pipeline is fully automated, and it supports optional human-in-the-loop refinement via visualized intermediate structures.

Our evaluation shows that \textsc{Req2LTL} substantially surpasses existing state-of-the-art methods, such as NL2SPEC~\cite{Cosler2023nl2specIT}, NL2LTL~\cite{Fuggitti2023NL2LTLA}, and NL2TL~\cite{Chen2023NL2TLTN}, on both academic benchmarks and real-world aerospace datasets. Specifically, \textsc{Req2LTL} achieves 88.4\% semantic accuracy and 100\% syntactic correctness, clearly demonstrating its efficacy in practical industrial contexts.
Ablation studies confirm the importance of hierarchical decomposition and rule-based validation in handling complex, nested logic.

Our contributions are summarized as follows:

\begin{itemize}
    \item\textbf{OnionL Intermediate Representation}: We introduce \textit{OnionL}, a hierarchical intermediate language explicitly encoding temporal semantics, scopes, and logical relations. \textit{OnionL} serves as a robust semantic bridge between natural language requirements and LTL formulas.
    \item\textbf{Hierarchical Semantic Decomposition}: We present a two-stage, prompt-driven decomposition algorithm (\textsc{Building-OnionL}) leveraging LLMs. This systematic approach significantly improves semantic accuracy and structural integrity in the translation process.
    \item\textbf{\textsc{Req2LTL} Framework}: We present \textsc{Req2LTL}, an automated framework combining LLM-based semantic decomposition (\textsc{Req2OnionL}) and deterministic rule-based translation (\textsc{OnionL2LTL}), supported by automated validation. It achieves superior accuracy (88.4\%) and perfect syntactic correctness (100\%) compared to existing approaches, across both academic and industrial benchmarks.
\end{itemize}

\section{Background and Motivation}
\begin{figure*}[t] 
  \centering 
  \includegraphics[width=1\textwidth]{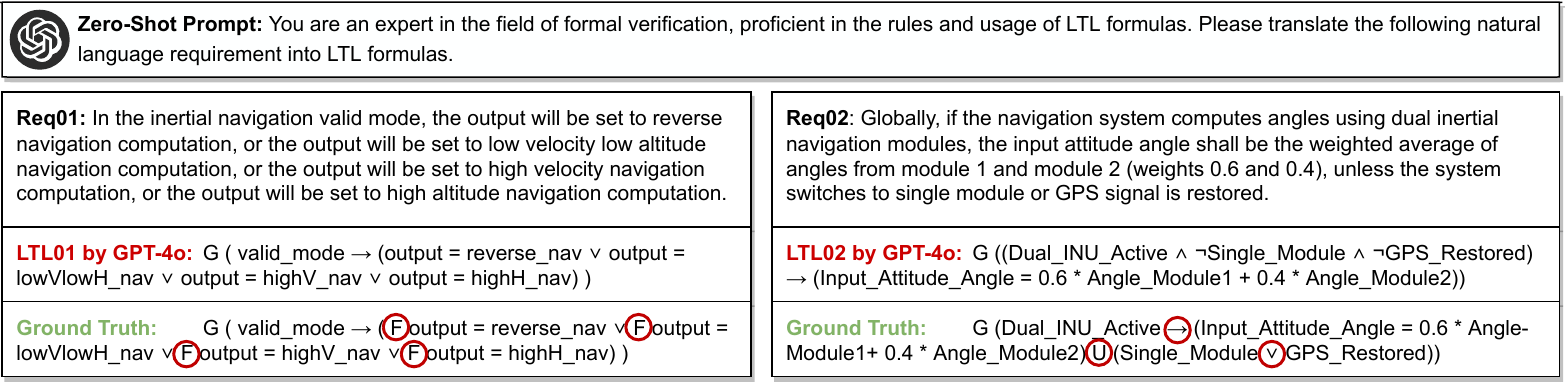}
  \setlength{\abovecaptionskip}{-15pt}
  \setlength{\belowcaptionskip}{0pt}
  \caption{\textbf{Examples showcasing the discrepancy between LLM generated LTL and ground truth for complex requirements.} The \textcolor{warmdarkred}{\textbf{red-circle}} indicates errors in the LLM's output due to missing details, contrasted with the manually corrected specification.}
  \label{fig:example of Space} 
\end{figure*}

\subsection{Background}
Linear Temporal Logic (LTL) is a formal modal logic extensively used to specify temporal properties over infinite execution paths. 
Formally, given a finite set of atomic propositions $\mathit{AP}$, LTL formulas follow the grammar:
\begin{equation*}
\varphi ::= p \;\vert\; \neg \varphi \;\vert\; \varphi_1 \land \varphi_2 \;\vert\; \mathbf{X} \varphi \;\vert\; \mathbf{F} \varphi \;\vert\; \mathbf{G} \varphi \;\vert\; \varphi_1 \;\mathbf{U}\; \varphi_2
\end{equation*}
where $p \in \mathit{AP}$. 
The temporal operators in LTL convey critical aspects of system dynamics: $\mathbf{X} \varphi$ specifies that $\varphi$ holds in the next state;  $\mathbf{F} \varphi$ indicates $\varphi$ eventually holds at some future state; $\mathbf{G} \varphi$ demands $\varphi$ to always hold; and $\varphi_1 \mathbf{U} \varphi_2$ requires $\varphi_1$ to hold $\varphi_2$ until becomes true.

Table~\ref{tab:nl_examples_ltl} provides common examples from academic benchmarks ~\cite{Chen2023NL2TLTN, Cosler2023nl2specIT, He2021DeepSTLF, Wang2020LearningAN, Fuggitti2023NL2LTLA}, where the natural language statements are typically concise, grammatically regular, and temporally explicit. 
The inclusion of keywords such as {\textit{always}} or {\textit{if...then}} allows for a direct correlation to standard LTL operators, facilitating a more regulated translation process.
In contrast, requirements derived from safety-critical industrial systems exhibit a significantly higher level of complexity. 
They are generally longer, more deeply nested, and enriched with domain-specific terminology. Often, these requirements extend across multiple sentences, incorporating implicit assumptions and composite logical constructs.

\subsection{Motivating Examples}

Figure~\ref{fig:example of Space} presents two representative real-world cases where GPT-4o's translation yields significant discrepancies from the correct LTL specifications.
The results in both cases are also consistent when employing  NL2SPEC~\cite{Cosler2023nl2specIT}.
\textbf{Req01} demonstrates LLMs' struggle with implicit temporal constraints. The requirement states that in ``\textit{inertial navigation valid mode},'' the system will eventually output one of four navigation computation types. 
While GPT-4o correctly maps the logical disjunction ({\small$\lor$}), it misses the eventuality constraint (\texttt{F} operator). The generated LTL {\small\texttt{G}$($\texttt{valid\_mode} $\rightarrow$ \texttt{output} $= ...)$} incorrectly implies that the output must be set immediately upon entering the mode. In contrast, the ground truth {\small\texttt{G}$($\texttt{valid\_mode} $\rightarrow$ \texttt{F} \texttt{output} $= ...)$} accurately reflects that the system may take a finite time to determine the appropriate computation—a nuance critical in aerospace systems.

Similarly, in \textbf{Req02}, the requirement involves dual inertial navigation logic; the {\textit{unless}} clause is misinterpreted as a simple negation rather than its intended meaning {\textit{until a condition occurs}}. 
GPT-4o's output introduces three critical flaws: (1) Missing implicit temporal cues: it omits the until ({\small\texttt{U}}) operator, failing to specify that the weighted average holds only until termination conditions occur.
(2) Flattened logical structure that loses nesting and scope boundaries: the ground truth {\small\texttt{G}$($\texttt{Dual\_INU\_Active} $\rightarrow$ $...$ \texttt{U} $($\texttt{Single\_Module} $\lor$ \texttt{GPS\_Restored}$)$} correctly prioritizes {\small\texttt{Dual\_INU\_Active}} as the sole trigger.
While the GPT-4o erroneously treats all conditions as co-occurring prerequisites, due to the flat logical structure fails to capture the hierarchical relationship:
{\small\texttt{G}$(...$$\rightarrow$$...$\texttt{U}$(...$$\lor$$...)$}.
This error stems from the inherent tendency of LLMs to prioritize syntactic translation over profound logical reasoning.
(3) Accumulation of errors and lack of validation: the formula produced by GPT-4o initially included {\small\texttt{Dual\_INU\_Active} $\land$ $\neg$\texttt{Single\_Module} $\land$ $\neg$\texttt{GPS\_Restored}}, which omitted the implies relationship, and subsequent generations were based on this deviation, resulting in further semantic deviation.
This mismatch leads to ``silent failures'', where the formula is syntactically valid but semantically incorrect, as highlighted by the red circles in Figure~\ref{fig:example of Space}.

To further investigate the limitations of using LLMs for translating natural language into LTL, we collected 112 natural language requirements from two real-world aerospace systems and conducted detailed empirical experiments. 
Under zero-shot prompting, GPT-4o generates only 49 semantically correct LTL formulas, highlighting a broader difficulty in capturing embedded temporal semantics when cues are implicit or structurally distant from their target propositions. 
However, the model demonstrates proficiency in extracting atomic propositions, achieving a recall rate of 98.5\%. 
These observations underscore a significant limitation of current LLM-based translation methods: while LLMs excel at extracting atomic propositions, they struggle with global logical synthesis.
In particular, LLMs are adept at recognizing isolated facts but lack the structured reasoning needed to correctly compose and validate nested temporal and logical relationships, especially when implicit cues are present.

\subsection{Towards a Two-stage, Structured Translation Approach}
To effectively overcome the above inherent limitations, a structured, hierarchical approach is crucial, combining the strengths of LLMs in local semantic interpretation and atomic proposition extraction with deterministic rule-based mechanisms for global logical synthesis. 

Motivated by these insights, we first introduce \textit{OnionL}, a hierarchical intermediate representation explicitly designed to bridge the semantic gap between natural language and LTL. 
By explicitly encoding temporal semantics and logical relations in a structured intermediate representation, we can effectively reduce the cognitive load on LLMs, allowing them to focus on accurate semantic parsing rather than complex logical composition.
Leveraging \textit{OnionL} representation, our proposed \textsc{Req2LTL} framework (Section~\ref{cha:3}) systematically decomposes requirements via the \textsc{Req2OnionL} module (Section~\ref{cha:4}).
It deterministically translates the resulting structured representations into accurate LTL formulas using the \textsc{OnionL2LTL} module (Section~\ref{cha:5}), thereby ensuring semantic accuracy and structural integrity.

\section{Global View of the Req2LTL Framework}\label{cha:3}
\begin{figure*}[!t] 
  \centering 
  \includegraphics[width=0.95\textwidth]{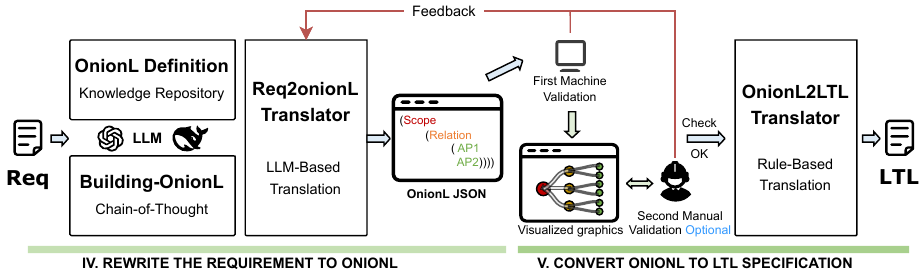}
  \setlength{\abovecaptionskip}{0pt}
  \setlength{\belowcaptionskip}{0pt}
  \caption{\textbf{Overview of our proposed \textsc{Req2LTL} framework pipeline.} Given a natural language requirement, the system incrementally constructs a compositional \textit{OnionL} tree via semantic prompting and structural parsing. The intermediate representation undergoes validation (machine and optional human-in-the-loop) and is then translated into an LTL formula by a rule-based engine. }
  \label{fig:overview} 
\end{figure*}

Figure~\ref{fig:overview} presents an overview of the \textsc{Req2LTL} framework, which enables the automatic translation of complex NL requirements into formal LTL specifications through a structured intermediate representation called \textit{OnionL}. 
Through staged modeling and a rule-based constraint mechanism, the framework ensures semantic consistency and structural correctness. 
The framework consists of two functionally independent yet complementary modules: \textsc{Req2onionL} and \textsc{OnionL2LTL}.

\subsection{Constructing the Structured ``Onion'' -- \textsc{Req2OnionL}}

The \textsc{Req2OnionL} (Section~\ref{cha:4}) module is responsible for transforming free-form NL requirements into structured \textit{OnionL} expressions. 
This module employs an LLM as its generative engine, with the generation process constrained and guided by a knowledge repository and chain of thought.

\textbf{Knowledge Repository} (Section~\ref{sec:4.1}). Within this repository, the syntactic structure and semantic composition of \textit{OnionL} are formally defined as a recursive linguistic system consisting of three types of semantic units: scopes, relations, and atomic propositions. The overall structure is mapped onto a tree-based syntactic framework.

\textbf{Chain-of-Thought} (Section~\ref{sec:4.2}). Our \textsc{Req2onionL} integrates a Chain-of-Thought prompting algorithm named \textsc{Building-OnionL}. This algorithm operates in two stages. The first stage performs macro-structural extraction, identifying global temporal or pattern-based scopes. The second stage conducts recursive decomposition and atomic proposition normalization, ultimately generating a logically complete and semantically faithful \textit{OnionL} JSON.

\subsection{Decomposing the Structured ``Onion'' -- \textsc{OnionL2LTL}}

The \textsc{OnionL2LTL} (Section~\ref{cha:5}) module takes the \textit{OnionL} JSON and translates it into a well-formed LTL formula through a combination of validation and rule-based synthesis.

\textbf{Dual Validation} (Section~\ref{5.1}). First, machine validation performs a depth-first traversal of the \textit{OnionL} tree, validating scope-clause pairing rules node by node. It checks the arity and type validity of unary and binary operators and identifies redundant chains, undefined operations, or logical conflicts. Second, for manual validation, considering potential sentence ambiguity and incomplete requirements in safety-critical scenarios, \textsc{OnionL2LTL} supports automatically rendering OnionL JSON into Mermaid~\cite{mermaidjs} formatted visual tree diagrams. This facilitates engineers in conducting semantic inspections efficiently, promoting a ``human-in-the-loop'' feedback and validation cycle.

\textbf{Rule-Based Translation} (Section~\ref{5.2}). Upon completing the validation, the \textsc{OnionL2LTL} module starts a rule-based translation into LTL. Each scope node is mapped to the corresponding unary operator, each relation node is converted into a binary operator, and each atomic proposition is reconstructed into a predicate expression according to its semantic label. This process is deterministic and structure-preserving, ensuring that each validated \textit{OnionL} JSON corresponds to a distinct LTL formula that is both syntactically and semantically correct.

\smallskip
Together, \textsc{Req2onionL} and \textsc{OnionL2LTL} form a pipeline that not only achieves high translation accuracy but also introduces transparency, modularity, and verifiability into the process of deriving LTL specifications from NL.


\section{Rewrite the Requirement to OnionL}
\label{cha:4}
\subsection{Design of the \textit{OnionL} Intermediate Language}
\label{sec:4.1}


\begin{tcolorbox}[
  colback=gray!5,
  colframe=black!30,
  boxrule=0.4pt,
  arc=2pt,
  left=4pt,
  right=4pt,
  top=3pt,
  bottom=3pt,
  fontupper=\itshape,
  before skip=10pt,
  after skip=10pt
]
``Onions have layers. You get it? We both have layers.'' \\
— \textit{Shrek (2001)}
\end{tcolorbox}

This humorous line captures the core idea behind \textit{OnionL}: natural language requirements, like onions, have layers. These layers represent nested semantic structures—temporal scopes, logical relations, and predicate-level facts—that must be preserved in formal specification.

Translating such requirements into LTL is inherently a structure-preserving task. 
While LTL uses a recursive grammar over logical and temporal operators, industrial requirements consistently follow \textbf{semantically layered patterns} due to strict demands on determinism, correctness, and timing.

To systematically capture these patterns, we introduce \textit{OnionL}, a hierarchical intermediate language that decomposes requirements into three compositional elements: \textit{scopes}, \textit{relations}, and \textit{atomic propositions}.
Each requirement is represented as a tree built over these constructs, enabling deterministic parsing and rule-based translation into LTL. \textit{OnionL} thus serves as a semantic bridge between informal natural language and formal temporal logic.

\noindent \textbf{Atomic Propositions (APs)} represent predicate-level facts about system behavior, such as sensor states, threshold conditions, or numeric assignments. Each AP is annotated with four semantic subfields to support symbolic interpretation:
\begin{itemize}
    \item \texttt{Com}: component or subsystem identifier;
    \item \texttt{Var}: variable name or symbolic constant;
    \item \texttt{Rel}: relational operator (\textit{e.g.}, $=$, $>$, $\leq$);
    \item \texttt{Formula}: numeric value or arithmetic expression.
\end{itemize}

These subfields are combined into canonical patterns, such as comparisons (\texttt{Var} $\texttt{Rel}$ \texttt{Formula}) and assignments (\texttt{Var} $=$ \texttt{Formula}).

\noindent \textbf{Scopes} define the contextual boundary in which a requirement holds. They correspond to unary operators in LTL and can be:
\begin{itemize}
    \item \textit{Temporal scopes}, \textit{e.g.}, {\small\texttt{Globally}}, {\small\texttt{Eventually}}, or {\small\texttt{Next}};
    \item \textit{Mode scopes}, which capture operational modes (\textit{e.g.}, \textit{``in navigation valid mode''}) and are always interpreted as antecedents in implications.
\end{itemize}

\noindent \textbf{Relations} define logical or temporal dependencies between subformulas. OnionL supports:
\begin{itemize}
    \item \textit{Logical relations}: conjunction ($\land$), disjunction ($\lor$), implication ($\rightarrow$);
    \item \textit{Temporal relations}: variants of the \texttt{Until} operator.
    \begin{itemize}
        \item \textit{Basic precedence} -- one condition eventually follows another;
        \item \textit{Sustained precedence} -- one condition must hold continuously until another occurs.
    \end{itemize}
\end{itemize}

\noindent \textbf{Recursive Grammar.}
Let $\mathit{AP}$ be the set of atomic propositions, $\mathit{SC}$ the set of unary scope operators, 
and $\mathit{RE}$ the set of binary relation operators. 
We denote $\varphi_{AP}\in \mathit{AP}$, $\varphi_{SC}\in \mathit{SC}$, and $\varphi_{RE}\in \mathit{RE}$ accordingly. 
An \textit{OnionL} expression $\varphi$ is defined recursively as:
{
    \begin{align*}
        &\text{(Base case)} \quad &&\varphi ::= \varphi_{\mathit{AP}} \in \mathit{AP} \\
        &\text{(Scope application)} \quad &&\varphi ::= \varphi_{\mathit{SC}}(\varphi_1) \\
        &\text{(Relational composition)} \quad &&\varphi ::= \varphi_1 \,\varphi_{\mathit{RE}}\, \varphi_2 \\
        &\text{(Nested combination)} \quad &&\varphi ::= \varphi_{\mathit{SC}}(\varphi_1) \,\varphi_{\mathit{RE}}\, \varphi_{\mathit{SC}}(\varphi_2)
    \end{align*}
} To avoid ambiguity, all relation operators are left-associative, and scopes bind more tightly than relations. 
For instance, the \textit{OnionL} tree for {\small $G(F(p) \lor q)$} would be placed {\small\texttt{Globally}} at the root and {\small\texttt{Eventually}} applied to $p$ as a nested child of the left operand.

\smallskip
\noindent \textbf{Illustrative Example.}
Consider the requirement: ``\textit{In valid mode, if the temperature exceeds 50, eventually the warning light is turned on.}''
Its corresponding \textit{OnionL} expression is:

{
\small
\begin{verbatim}
(Globally
    (Implies
        (Atomic: "workmode = valid")
        (Eventually
            (Implies
                (Atomic: "temperature > 50")
                (Atomic: "warning = ON")))))
\end{verbatim}
}

This structure reflects the following semantic layering.
A top-level {\small\texttt{Globally}} scope encapsulates the entire implication;
The antecedent is a condition based on the system working mode;
The consequent contains a nested implication under the \texttt{Eventually} operator.
Such structured representation ensures that both temporal and conditional semantics are preserved and faithfully encoded in a compositional form. 
By compositionality we mean that LTL has an inductively defined syntax where complex formulas are built from subformulas via unary or binary operators, and its semantics is defined per operator. Hence, the meaning of a formula is determined compositionally from the meanings of its subformulas.

\subsection{Two-Stage Decomposition Algorithm}
\label{sec:4.2}

We propose \textsc{Building-OnionL}, a hierarchical prompt-driven algorithm that transforms natural language requirements into structured \textit{OnionL} trees. The algorithm operates in two semantic stages. \textbf{Stage I}: Macro-Structure Extraction. Identify the global semantic scope (Temporal or Mode) and construct the top-level \textit{OnionL} node.
\textbf{Stage II}: Recursive Clause Decomposition. Decompose the remaining clause into nested scopes, relations, and atomic propositions until a fully structured tree is obtained.
This staged strategy mirrors the compositional form of LTL and enforces semantic anchoring prior to structural expansion, enabling faithful alignment between requirement semantics and formal representations.
Algorithm~\ref{alg:building_OnionL_full} details its formal pseudo code, which comprises two stages and six steps.

\textbf{Step 1:} Scope Identification and Clause Separation.
The model first identifies the overarching semantic scope of the requirement. This includes:
Temporal scopes and Mode scopes. If no explicit marker is present, a default {\small\texttt{Globally}} scope is assumed.
The remaining part of the sentence is separated as the main clause to be recursively processed.

\textbf{Step 2:} Scope Type Analysis and Top-Level Construction.
If a temporal scope is detected, it is assigned as the root of the \textit{OnionL} tree, with the remaining clause as its child. If the clause is atomic, the algorithm skips recursive decomposition and proceeds to Step~6.  
For mode scopes, the scope is interpreted as a logical antecedent, and the root is set to {\small\texttt{Globally}}, forming an implication from the mode condition to the consequent behavior. If no scope is found, the main clause is wrapped by {\small\texttt{Globally}} by default.

\textbf{Step 3:} Unary Operator Identification and Extraction.
The model examines whether the current clause contains a unary temporal operator. If found, this operator becomes a new parent node, and the remaining clause is recursively parsed as its child node.

\textbf{Step 4:} Binary Logical Operator Decomposition.
If the clause contains a binary logical connective, the clause is split into two subtrees, with the operator as the parent node. Each part is recursively processed.

\textbf{Step 5:} Atomicity Judgment and Recursive Termination.
If neither unary nor binary operators are detected, the model evaluates whether the clause is atomic. If it is, the clause is encapsulated as a leaf node. Otherwise, the model devises a refined decomposition plan and reapplies Steps 3 through 5 to the revised clause.

\textbf{Step 6:} Semantic Reduction and AP Normalization.
Once a clause is confirmed to be atomic, the final step performs semantic normalization. The clause is rewritten into one of the five predefined atomic proposition patterns, explicitly populating the \texttt{Com}, \texttt{Var}, \texttt{Rel}, and \texttt{Formula} fields. This ensures machine-checkable alignment with the engineering semantics expected by downstream verification tools.

\setlength{\textfloatsep}{4pt}
\begin{algorithm}[t]
\algsetup{linenosize=\footnotesize} \footnotesize
\caption{\textsc{Building-OnionL}}
\label{alg:building_OnionL_full}
\begin{minipage}{\linewidth}
\KwIn{Natural language requirement $\mathcal{R}$}%
\KwOut{Structured \textit{OnionL} expression $\varphi$}%
\textcolor{gray}{\textit{\small// Stage I: Macro-Structure Extraction}}\\
\textcolor{gray}{\noindent\textit{\small// Step 1: Scope Identification and Clause Separation}}\\
$\mathsf{scope},\, \mathsf{clause} \gets \textsc{ExtractScope}(\mathcal{R})$\;
\eIf{$\mathsf{scope}$ is Temporal}{
  \textcolor{gray}{\small\textit{// Step 2: Temporal scope $\rightarrow$ root node}}\\
  \If{\textsc{IsAtomicClause}($\mathsf{clause}$)}{
    \textcolor{gray}{\small\textit{// Step 6: Atomic clause, normalize and terminate}}\\
    \Return \textsc{NormalizeAP}($\mathsf{clause}$)
  }
  $\varphi \gets \mathsf{scope}(\textsc{DecomposeClause}(\mathsf{clause}))$\;
}{
  \If{$\mathsf{scope}$ is Mode-based}{
    \textcolor{gray}{\small\textit{// Step 2: Mode scope $\rightarrow$ G(antecedent $\rightarrow$ consequent)}}\\
    $\varphi \gets \mathbf{G}(\mathsf{scope} \rightarrow \textsc{DecomposeClause}(\mathsf{clause}))$\;
  }
  \Else{
    \textcolor{gray}{\textit{\small// Step 2 (fallback): Wrap clause with $G$}}\\
    $\varphi \gets \mathbf{G}(\textsc{DecomposeClause}(\mathsf{clause}))$\;
  }
}
\textcolor{gray}{\small\textit{// Stage II: Recursive Clause Decomposition and AP Normalization}}\\
\SetKwFunction{FDecompose}{DecomposeClause}
\SetKwProg{Fn}{Function}{:}{}
\Fn{\FDecompose{$c$}}{
  \If{\textsc{HasUnaryScope}($c$)}{
    \textcolor{gray}{\small\textit{// Step 3: Unary scope $\rightarrow$ parent}}\\
    $\mathsf{scope},\, c' \gets \textsc{ExtractUnaryScope}(c)$\;
    \Return $\mathsf{scope}(\FDecompose(c'))$
  }
  \If{\textsc{HasBinaryRelation}($c$)}{
    \textcolor{gray}{\small\textit{// Step 4: Binary relation $\rightarrow$ subtrees}}\\
    $c_1, c_2, \mathsf{rel} \gets \textsc{ExtractBinaryRelation}(c)$\;
    \Return $\FDecompose(c_1)\, \mathsf{rel}\, \FDecompose(c_2)$
  }
  \If{\textsc{IsAtomicClause}($c$)}{
    \Return \textsc{NormalizeAP}($c$)\textcolor{gray}{\small\textit{// Step 5: Atomic clause, stop}}
  }
  $\mathsf{refined} \gets \textsc{RefineImplicitStructure}(c)$\;
  \Return \FDecompose($\mathsf{refined}$)\textcolor{gray}{\small\textit{// Step 5 (fallback): Refine}}
}
\end{minipage}%
\end{algorithm}

\begin{figure*}[!t] 
  \centering 
  \includegraphics[width=1.025\textwidth]{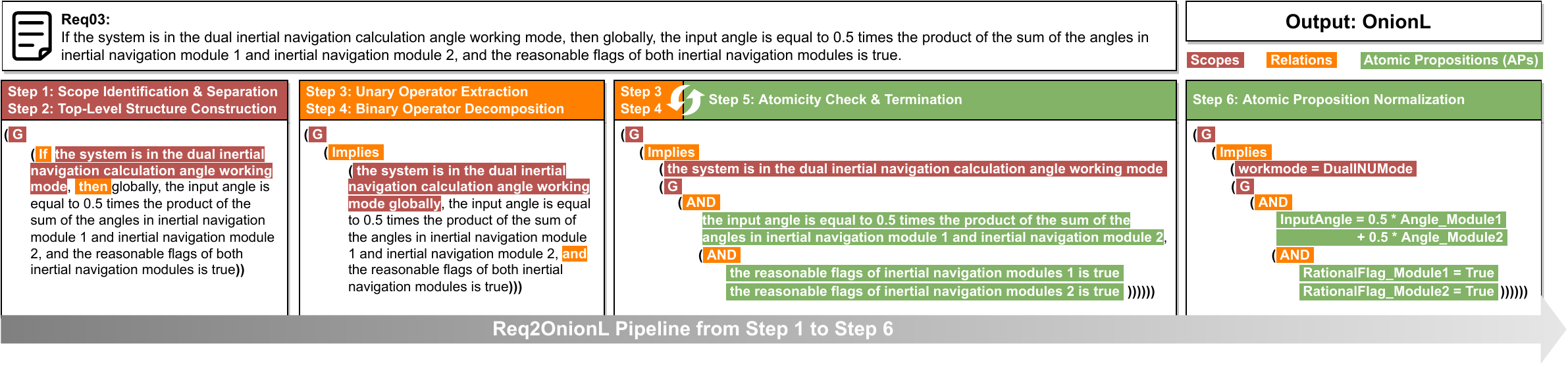}
  \vspace{-1cm}
  \caption{End-to-end transformation of a natural language requirement into a structured \textit{OnionL} tree by the \textsc{Req2onionL} translator. The figure illustrates the full execution trace of the \textsc{Building-onionL} algorithm, from top-level scope extraction to atomic proposition normalization. The resulting \textit{OnionL} structure captures the hierarchical semantics of the input and enables deterministic translation into formal LTL.}
  \label{fig:req2onionl_example} 
\end{figure*}
\subsection{The Req2OnionL Translator}
\label{sec:4.3}

The \textsc{Req2OnionL} translator forms the front-end of the \textsc{Req2LTL} framework. It transforms natural language requirements into structured \textit{OnionL} trees by prompting LLMs under formal guidance. We divide this process into two key components.
We treat the \textit{OnionL} representation introduced in Section~\ref{sec:4.1} as a formal knowledge base. It defines the compositional syntax and semantic elements that serve as structural priors to constrain the generation space and ensure syntactic correctness.
We utilize the \textsc{Building-OnionL} algorithm, introduced in Section~\ref{sec:4.2}, as a chain-of-thought prompting strategy. It guides the model to perform structured reasoning in two stages: first identifying the global scope, then recursively decomposing the sentence into a semantically faithful tree.
To further improve robustness, we introduce grammar-constrained prompt templates along with a few task-specific few-shot examples. These examples capture common patterns in industrial requirements—such as mode-triggered behavior, temporal constraints, and threshold logic—and help the model generalize across structural variations.
The output is a machine-readable \textit{OnionL} JSON object that preserves the hierarchical semantics of the requirement. This serves as input to downstream validation and formal translation. Figure~\ref{fig:req2onionl_example} shows an example of the end-to-end result. As illustrated, the construction mirrors Algorithm~1 step by step: 
(1) detecting the global scope, 
(2) building the root and separating the main clause, 
(3) extracting unary operators, 
(4) splitting clauses by binary relations, 
(5) performing atomicity checks, and 
(6) normalizing atomic propositions. 
The right panel shows the resulting \textit{OnionL} JSON aligned with the original requirement.

\section{Convert onionL to LTL specification}
\label{cha:5}


\subsection{Validation of the OnionL}
\label{5.1}


\textbf{First Machine Validation.}
The system performs a recursive, depth-first traversal of the \textit{OnionL} to verify structural well-formedness. 
Each node is checked against its expected role in LTL: unary operators must have exactly one valid child, binary operators must have two, and atomic propositions must occur only at the leaves.
In addition, the system checks for \textit{OnionL}-specific composition constraints. 
Atomic propositions are expected to contain one or more well-formed subfields—such as \texttt{Com}, \texttt{Var}, \texttt{Rel}, and \texttt{Formula}—depending on the semantics of the requirement.
The validator checks for compatibility among these subfields and flags missing or contradictory combinations where applicable; scope nesting is validated against a finite set of legal patterns. 
The validator detects redundant or malformed constructs, such as deeply nested \texttt{AND} chains, and rewrites them into canonical left-associative forms. 
Violations are reported with detailed diagnostic logs, including error type, tree path, and suggested fixes.

\begin{figure}[t]
\centering
\includegraphics[width=1\columnwidth]{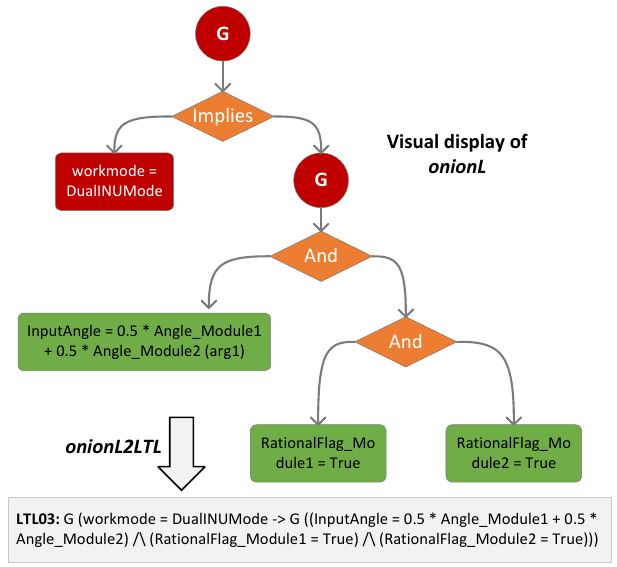}
\setlength{\abovecaptionskip}{-2.5pt}
\setlength{\belowcaptionskip}{0pt}
\vspace{-0.3cm}
\caption{Visualized \textit{OnionL} structure for human-in-the-loop validation.
The diagram presents the model’s inferred semantics in a structured form, making it easier for engineers to identify semantic errors. Once validated, it is deterministically translated into an LTL formula via rule-based synthesis.}
\label{fig:onionl-visual}
\end{figure}

\textbf{Second Manual Validation.}
For safety-critical requirements, human-in-the-loop validation is strongly recommended. The \textit{OnionL} structure is visualized as a directed graph using Mermaid~\cite{mermaidjs} syntax, as shown in Figure~\ref{fig:onionl-visual}. This graphical representation exposes the hierarchical semantics inferred by the model in a readable and structured format, allowing engineers to inspect scopes, logical relationships, and atomic conditions efficiently. When semantic inconsistencies or structural errors are identified, engineers can directly modify the \textit{OnionL} tree or provide feedback to the system, prompting the \textsc{Req2onionL} module to regenerate a corrected structure. Once verified, the validated \textit{OnionL} is passed to the rule-based translator for deterministic LTL synthesis.

\subsection{Rule-Based Translation to LTL}
\label{5.2}

Once the \textit{OnionL} structure passes validation, it is forwarded to the \textsc{OnionL2LTL} translator—a dedicated module responsible for converting structured trees into standard LTL formulas. This module implements a set of fixed, rule-based mappings that deterministically translate each semantic element of the tree into its corresponding LTL syntax. Formally, we define a total function $T: \textit{OnionL} \rightarrow \textit{LTL}$ that maps each node in the tree to its corresponding logical form.

The translation proceeds recursively in a post-order traversal of the \textit{OnionL} tree. For each node, the algorithm first processes all of its children before applying the corresponding transformation at the parent level. For unary nodes such as {\small\texttt{Globally}}, {\small\texttt{Eventually}}, or {\small\texttt{Next}}, the child node is fully translated first, after which the appropriate temporal operator is applied. For binary nodes such as {\small\texttt{AND}}, {\small\texttt{OR}}, or {\small\texttt{Implies}}, both child nodes are recursively translated before being combined using logical connectives such as $\wedge$, $\vee$, or $\rightarrow$. For atomic propositions at the leaves, the expression is directly reconstructed from the subfields \texttt{Com}, \texttt{Var}, \texttt{Rel}, and \texttt{Formula}, yielding a flattened predicate representation.

All translation rules in the \textsc{OnionL2LTL} module are deterministic and structure-preserving: 
each well-formed \textit{OnionL} tree corresponds to exactly one valid LTL formula. 
Provided that the \textit{OnionL} representation faithfully captures the semantics of the original requirement, the resulting LTL is guaranteed to be both syntactically correct and semantically consistent. 
This rule-driven translation approach enables seamless integration with existing model checkers such as NuSMV~\cite{Cimatti2000NuSMV}, while Spot~\cite{DuretLutz2016Spot2} is used as an LTL/$\omega$-automata manipulation and parsing framework.
It also supports automated batch conversion of requirements into formal specifications, making the framework scalable and robust for industrial verification workflows.

\section{Evaluation}
\label{cha:6}
To assess the effectiveness of the proposed \textsc{Req2LTL} framework, we conduct a comprehensive evaluation aimed at answering the following research questions:
\begin{steps}\setlength{\itemsep} {0.15cm}
\item How does \textsc{Req2LTL} perform on standard academic benchmarks compared to existing state-of-the-art NL-to-LTL translation methods?
\item To what extent can \textsc{Req2LTL} accurately and reliably translate real-world industrial requirements into semantically correct and syntactically valid LTL specifications?
\item What is the contribution of the \textit{OnionL} structure, the \textsc{Building-OnionL} algorithm, and the validation feedback mechanism to the overall performance?
\item Are structural challenges in industrial requirements (\textit{e.g.}, temporal ambiguity, nested logic) systematic and significant? How effectively does \textsc{Req2LTL} address and mitigate these complexities?
\end{steps}

\subsection{Experimental Setup}

\subsubsection{Datasets} \label{sec:6.1}

We evaluate two categories of data.

\textit{Academic Benchmark:}
To assess cross-domain generalization, we evaluate \textsc{Req2LTL} on a public benchmark comprising three curated subsets: \textit{Circuit}, \textit{Navigation}, and \textit{Office Email}, originally proposed in prior work~\cite{Chen2023NL2TLTN, He2021DeepSTLF, Wang2020LearningAN, Fuggitti2023NL2LTLA}. These requirements are shorter, less ambiguous, and syntactically regular, typically mapping directly to standard LTL constructs.
The benchmark adopts a \emph{lifted} representation in which all atomic propositions are abstracted into placeholders (\textit{e.g.}, \texttt{Prop1}, \texttt{Prop2}). \textsc{Req2LTL} accommodates this abstraction format by applying placeholder substitution during its \textsc{OnionL2LTL} post-processing stage.

\textit{Industrial Dataset:}
Our dataset is built from requirement documents provided by our industrial aerospace partners. These requirements originate from two critical spacecraft systems: a \textit{sun-search controller} system, which adjusts solar panels or spacecraft attitude to track the Sun, and a \textit{propulsion management} system, which governs thruster activation and fuel flow during orbital maneuvers and attitude adjustments. 
In total, we collect 112 requirements covering diverse operational scenarios, including initialization, attitude determination, anomaly handling, and fault tolerance. 
Compared to academic datasets, these industrial specifications are characterized by longer sentence lengths (avg. 43.7 tokens), deeper logical nesting (over 60\% with $\geq$ 2 layers), and domain-specific vocabulary grounded in hardware and control logic. 
A detailed structural complexity analysis is presented in Section~\ref{sec:structural-complexity-analysis}.
To ensure high-quality annotations, each requirement was independently translated into an LTL formula by two annotators. 
These annotators were selected from a team of five experts, including two aerospace engineers and three formal verification researchers. 
Discrepancies between the two annotations were resolved through expert discussion, and all finalized formulas were reviewed by a senior engineer to ensure semantic fidelity and consistency. 
During this process, we did encounter certain requirements that either involved explicit quantitative timing (e.g., response within 5 seconds) or lacked any temporal constraint. 
Since such requirements fall outside the expressive power of LTL, they were excluded from our benchmark to maintain semantic compatibility. 
The resulting set of 112 validated Req–LTL pairs constitutes our gold dataset and serves as the reference standard for evaluation throughout this study.

\subsubsection{Baselines}
We compare \textsc{Req2LTL} with several baseline methods across two dimensions: model backend and generation strategy.
For model backends, we choose two recent and representative large language models: {GPT-4o}~\cite{openai2024gpt4o} and {DeepSeek-V3}~\cite{zhang2024deepseek}.
For generation strategies, we consider four methods with increasing levels of structural guidance. 
\textit{Zero-Shot Prompt} represents a direct instruction method that asks the model to translate a requirement into LTL without templates or intermediate representations.
{NL2LTL}~\cite{Fuggitti2023NL2LTLA} is a Python package developed by IBM Research that uses LLMs to translate NL instructions into LTL formulas.
{NL2SPEC}~\cite{Cosler2023nl2specIT} adopts a template-guided prompting approach that builds LTL expressions in stages, emphasizing interpretability and semantic traceability. 
{NL2TL}~\cite{Chen2023NL2TLTN}, originally designed for lifted STL translation, is adapted here as a prompt-based baseline for abstracted NL-to-LTL generation. 
We retain its prompt structure and abstraction convention, applying position-consistent placeholders to both inputs and references.

\begin{table}[t]
\centering
\setlength{\abovecaptionskip}{0pt}
\setlength{\belowcaptionskip}{0pt}
\caption{Performance Comparison on Academic Benchmarks (Lifted)}
\label{tab:cross_dataset_nl2tl}
\begin{tabular}{llcc}
\toprule
\textbf{Method} & \textbf{Domain} & \textbf{Binary Acc. (\%)} & \textbf{BLEU Score} \\
\midrule

\rowcolor{gray!10}
& Circuit      & 95.30 & \textbf{0.97} \\
\rowcolor{gray!10}
\textbf{Req2LTL}   & Navigation   & 94.50 & \textbf{0.97} \\
\rowcolor{gray!10}
                & Office Email & \textbf{96.70} & \textbf{0.98} \\

\midrule
\rowcolor{blue!5}
           & Circuit      & \textbf{96.20} & 0.96 \\
\rowcolor{blue!5}
NL2TL            & Navigation   & \textbf{96.50} & \textbf{0.97} \\
\rowcolor{blue!5}
                & Office Email & 96.00 & 0.96 \\

\midrule
\rowcolor{green!5}
         & Circuit      & 90.50 & 0.93 \\
\rowcolor{green!5}
 NL2LTL                & Navigation   & 89.90 & 0.93 \\
\rowcolor{green!5}
                & Office Email & 90.80 & 0.93 \\

\midrule
\rowcolor{orange!5}
        & Circuit      & 90.80 & 0.93 \\
\rowcolor{orange!5}
   NL2SPEC              & Navigation   & 89.70 & 0.93 \\
\rowcolor{orange!5}
                & Office Email & 91.20 & 0.93 \\

\bottomrule
\end{tabular}
\end{table}

\subsubsection{Evaluation Metrics}
\label{sec:6.3}

To comprehensively evaluate performance, we employ a multi-metric framework that integrates automated tools with expert review, based on:
(1) \textit{LTL Syntax Validity:} assesses whether the generated formula adheres to formal syntax. We use Spot to parse each formula; a formula is valid if it can be compiled into a Büchi automaton.
(2) \textit{Exact Match Accuracy:} measures semantic and logical equivalence to the reference. Two experts independently assess each pair, and a match is accepted only upon agreement.
(3) \textit{Atomic Proposition Recall:} evaluates coverage of key atomic propositions by comparing the AP sets of the output and the gold standard.
(4) \textit{BLEU Score:} estimates token-level structural similarity. To reduce lexical bias, atomic propositions are abstracted before scoring, focusing the metric on syntactic alignment.
All methods are tested on the same dataset under consistent procedures to ensure fairness. 
Importantly, all results reported in RQ1 and RQ2 were obtained from a fully automated pipeline, 
with the optional manual validation feature disabled. This ensures that the reported performance metrics reflect the framework’s automated capability without any human intervention.

\begin{table*}[t]
\centering
\setlength{\abovecaptionskip}{0pt}
\setlength{\belowcaptionskip}{0pt}
\caption{Comparison of Prompting Strategies and Ablation Study on \textsc{Req2LTL} Framework. \textbf{Bold} and \underline{underlined} values indicate the best and second-best results, respectively.}
\begin{tabular}{lcccc}
\toprule
\textbf{Setting} & \textbf{Exact Match (\%)} & \textbf{Syntax Validity (\%)} & \textbf{AP Recall (\%)} & \textbf{BLEU Score} \\
\midrule
\multicolumn{5}{l}{\textbf{Prompt Strategy Comparison}} \\
GPT-4o + Zero-Shot         & 43.8 & 89.3  & 98.5 & 0.76 \\
GPT-4o + NL2LTL            & 55.4 & 91.5  & 98.4 & 0.77 \\
GPT-4o + NL2SPEC           & 56.3 & 91.9  & 98.5 & 0.78 \\
GPT-4o + NL2TL             & 65.2 & 94.6  & 98.1 & 0.85 \\ 
\rowcolor{gray!10} 
\textbf{GPT-4o + Req2LTL}  & \textbf{88.4} & \textbf{100.0} & \textbf{99.5} & \textbf{0.96} \\ 
\midrule
DeepSeek-V3 + Zero-Shot         & 34.8 & 87.5  & 97.5 & 0.73 \\
DeepSeek-V3 + NL2LTL            & 54.0 & 91.2  & 98.6 & 0.78 \\
DeepSeek-V3 + NL2SPEC           & 54.5 & 91.9  & 98.7 & 0.79 \\
DeepSeek-V3 + NL2TL             & 58.0 & 92.9  & 99.1 & 0.82 \\
\rowcolor{gray!10} 
\textbf{DeepSeek-V3 + Req2LTL} & \underline{86.6} & \textbf{100.0} & \underline{99.2} & \textbf{0.96} \\
\midrule
\multicolumn{5}{l}{\textbf{Ablation Study on Req2LTL}} \\
\rowcolor{gray!10} \textbf{D Full Version}         & \textbf{88.4} & \textbf{100.0} & \textbf{99.5} & \textbf{0.96} \\
A w/o Structured OnionL        & 65.2 & 98.2  & 98.9 & 0.86 \\
B w/o Stage-wise Decomposition & 58.9 & 94.6  & 98.1 & 0.85 \\
C w/o Verification Feedback    & 84.8 & 90.2  & \textbf{99.5} & 0.91 \\
\bottomrule
\end{tabular}
\label{tab:prompt_ablation_final}
\end{table*}

\subsection{RQ1. Performance on Academic Benchmarks}

We evaluate the generalization performance of \textsc{Req2LTL} on three widely used academic benchmarks—\textit{Circuit}, \textit{Navigation}, and \textit{Office Email}—using GPT-4o as the backend.
Two automatic metrics are used: \textit{Binary Accuracy} measures exact string-level matches between generated and reference formulas without normalization or semantic equivalence checking, ensuring a strict evaluation of structural correctness. \textit{BLEU Score}, in contrast, offers a softer measure of syntactic similarity by evaluating token-level overlap, emphasizing partial alignment and structural fluency.
For academic benchmarks, we report only binary accuracy and \textit{BLEU} because the formulas are short, 
syntactically regular, and use abstract placeholders, 
making \textit{AP Recall} less informative. Moreover, binary accuracy already reflects both syntactic and emantic correctness in these settings.

As shown in Table~\ref{tab:cross_dataset_nl2tl}, \textsc{Req2LTL} achieves strong and consistent performance across all three domains.
It attains Binary Accuracy scores of 95.3\%, 94.5\%, and 96.7\% respectively, along with BLEU scores $\geq$ 0.97—higher than all other methods.
These results confirm that our hierarchical, two-stage translation framework generalizes effectively even in less ambiguous settings.
Notably, its performance remains steady despite variations in structure and content across datasets. Furthermore, our comparison of \textsc{Req2LTL} with three representative baselines demonstrates superior performance in BLEU scores comprehensively.
NL2TL achieves comparable binary accuracy (96.0–96.5\%) but exhibits slightly lower BLEU scores, suggesting minor structural drift.
NL2SPEC and NL2LTL perform similarly in both metrics, with binary accuracy around 90\% and BLEU scores near 0.93, reflecting moderate semantic precision but weaker syntactic control compared to \textsc{Req2LTL} and NL2TL.

\smallskip
\noindent\textbf{The answer to RQ1:} 
\textsc{Req2LTL} These results indicate that \textsc{Req2LTL} not only retains high precision in simplified formalization tasks but also outperforms existing prompting baselines in syntactic fidelity. Compared to other existing methods, its modular architecture and intermediate representation (\textit{OnionL}) allow robust alignment between natural language and logic, even in abstracted settings.
On academic benchmarks, \textsc{Req2LTL} performs comparably to NL2TL; our primary gains arise on semantically complex industrial requirements.

\subsection{RQ2: Performance on Industrial Requirements}

We evaluate the effectiveness of \textsc{Req2LTL} on a real-world industrial dataset comprising 112 requirements from aerospace control systems. This dataset features domain-specific terminology, complex conditional logic, and implicit temporal semantics, posing significantly greater challenges than academic benchmarks.

As shown in Table~\ref{tab:prompt_ablation_final}, \textsc{Req2LTL} substantially outperforms all baseline prompting methods across both models.
Under GPT-4o, it achieves an exact match rate of \textbf{88.4\%}, full \textbf{100.0\%} syntax validity, and a BLEU score of \textbf{0.96}, far surpassing the closest alternative (NL2TL at 65.2\% exact match and 0.85 BLEU). Similarly, when paired with DeepSeek-V3, \textsc{Req2LTL} maintains strong performance with \underline{86.6\%} exact match and identical BLEU (0.96), indicating consistent robustness across LLM backends.

Zero-shot prompting yields poor exact match rates (43.8\% with GPT-4o and 34.8\% with DeepSeek-V3), highlighting the inadequacy of unguided LLMs in handling complex industrial semantics. 
While NL2SPEC, NL2LTL, and NL2TL improve accuracy to the 55--65\% range, they still struggle with nested logic, implicit temporal cues, and domain-specific phrasing. 
Despite enhancements, their syntax validation remains below perfect ($<$95\%), while \textsc{Req2LTL} maintains superior syntax validity at 100\%, credited to its rule-based translation engine, which ensures structural correctness by construction.
Moreover, its high AP Recall (99.5\%) confirms that atomic propositions are effectively extracted, while the gap between recall and exact match reflects the model’s ability to preserve both semantics and structure.

\smallskip
\noindent\textbf{The answer to RQ2:} 
\textsc{Req2LTL} achieves best performance on complex industrial requirements, outperforming state-of-the-art methods by a wide margin. 
Utilizing a hierarchical \textit{OnionL} intermediate representation, it adeptly addresses significant obstacles like temporal ambiguity, logical nesting, and error propagation.
The integration of prompt-guided semantic decomposition and deterministic synthesis results in generating high-fidelity LTL formulas with complete syntactic accuracy.

\begin{table}[t]
\centering
\setlength{\abovecaptionskip}{0pt}
\setlength{\belowcaptionskip}{0pt}
\caption{Structural Complexity Comparison: Industrial vs. Academic}
\label{tab:structural-complexity}
\begin{tabular}{lcc}
\toprule
\textbf{Metric} & \textbf{Industrial} & \textbf{Academic} \\
\midrule
Avg. sentence length (tokens) & 43.7 & 15.3 \\
With $\geq$2-layer nesting & 63.2\% & 9.8\% \\
Avg. number of APs            & 3.7 & 2.1 \\
\bottomrule
\end{tabular}
\end{table}

\subsection{RQ3. Ablation Study}

To assess the individual contributions of key components within the \textsc{Req2LTL} framework, we conduct an ablation study using GPT-4o as the backend model.
Specifically, we compare four configurations: \textbf{D} (Full version): Includes structured modeling \textit{OnionL}, staged prompting, and verification feedback. \textbf{A} (No Structured Modeling): Bypasses OnionL construction; the LLM generates LTL directly from raw natural language. \textbf{B} (No Staged Prompting): Uses OnionL but removes step-wise construction, replacing it with a single monolithic prompt. \textbf{C} (No Verification Feedback): Retains modeling and prompting, but skips structural review before translation.

As shown in Table~\ref{tab:prompt_ablation_final}, the full version (D) consistently outperforms all ablated variants. Removing structured modeling (A) causes semantic accuracy to drop from 88.4\% to 65.2\%, and BLEU score decreases by more than 10 points. Eliminating staged prompting (B) leads to the lowest semantic accuracy (58.9\%), showing that incremental construction is essential for preserving logical structure. Removing verification feedback (C) primarily affects syntactic validity, which drops by nearly 10\%, suggesting its role in ensuring output correctness.
These results indicate that the observed $>20\%$ performance gain 
(from Configuration~A to the full version~D) stems from the 
combined effect of structured modeling via \textit{OnionL} 
and stage-wise decomposition, rather than from OnionL alone. 
Removing either component leads to significant degradation, 
showing that both are indispensable and mutually reinforcing.

\smallskip
\noindent\textbf{The answer to RQ3:}
Each core component of the \textsc{Req2LTL} pipeline contributes substantially to final performance. 
Structured intermediate representation (\textit{OnionL}) captures semantic hierarchy, staged prompting stabilizes generation, and verification feedback ensures correctness. 
Together, these modules enable \textsc{Req2LTL} to outperform existing approaches and deliver high-quality translations.

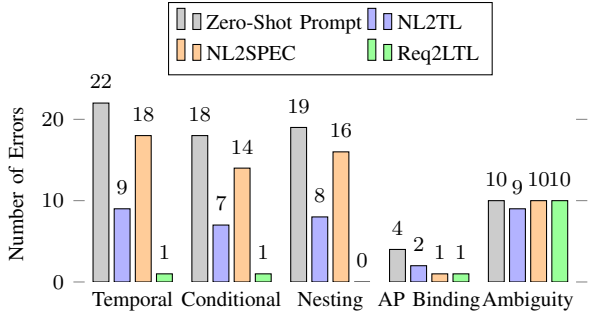
\begin{figure}[t]
\centering
\begin{tikzpicture}
\begin{axis}[
    ybar,
    bar width=6pt,
    width=0.95\columnwidth,
    height=4.2cm,
    ymin=0,
    enlarge x limits=0.15,
    ylabel={Number of Errors},
    symbolic x coords={
        Temporal, Conditional, Nesting, AP Binding, Ambiguity
    },
    xtick=data,
    xtick align=inside,
    xtick style={draw=none},
    every xtick label/.style={font=\tiny, rotate=25, anchor=east},
    ylabel style={font=\small},
    tick label style={font=\footnotesize},
    label style={font=\footnotesize},
    legend style={
        font=\footnotesize,
        cells={anchor=west},
        at={(0.5,1.05)},
        anchor=south,
        legend columns=2
    },
    nodes near coords,
    nodes near coords align={center},
    every node near coord/.append style={
        font=\footnotesize,
        yshift=2pt,
        anchor=south
    },
    grid=none,
    axis line style={draw=none}
]

\addplot[fill=gray!40] coordinates {
    (Temporal,22) 
    (Conditional,18) 
    (Nesting,19) 
    (AP Binding,4) 
    (Ambiguity,10)
};

\addplot[fill=blue!30] coordinates {
    (Temporal,9) 
    (Conditional,7) 
    (Nesting,8) 
    (AP Binding,2) 
    (Ambiguity,9)
};

\addplot[fill=orange!40] coordinates {
    (Temporal,18) 
    (Conditional,14) 
    (Nesting,16) 
    (AP Binding,1) 
    (Ambiguity,10)
};

\addplot[fill=green!40] coordinates {
    (Temporal,1) 
    (Conditional,1) 
    (Nesting,0) 
    (AP Binding,1) 
    (Ambiguity,10)
};

\legend{Zero-Shot Prompt, NL2TL, NL2SPEC, Req2LTL}
\end{axis}
\end{tikzpicture}
\setlength{\abovecaptionskip}{-7.5pt}
\setlength{\belowcaptionskip}{-12.5pt}

\caption{Error distribution across five categories for four methods. \textsc{Req2LTL} significantly reduces structural errors, with remaining issues primarily due to ambiguity or contextual omissions.}
\label{fig:error-types}
\end{figure}

\subsection{RQ4. Structural Complexity and Error Analysis}
\label{sec:structural-complexity-analysis}

To investigate whether structural challenges in industrial requirements are systematic and whether \textsc{Req2LTL} mitigates them, we conduct both quantitative and qualitative analyses. Table~\ref{tab:structural-complexity} compares our industrial aerospace dataset with standard academic benchmarks.
Industrial requirements are significantly more complex across all measured dimensions: they are longer on average, more deeply nested, and reference more atomic propositions than academic samples.
These features substantially increase both syntactic complexity and semantic ambiguity, posing challenges for LLMs.

To investigate how these complexities impact model performance, we classify all incorrect LTL outputs into five categories of structural errors: 
(1) Temporal Misinterpretation: Failure to capture correct temporal intent, such as precedence or sustained conditions. 
(2) Conditional Confusion: Logical inversion or misinterpretation of constructs like \textit{if} and \textit{unless}.
(3) Loss of Nesting: Flattening of multi-layer logical structures, losing semantic hierarchy.
(4) Incorrect AP Binding: Errors in identifying variables, thresholds, or assignments.
(5) Ambiguity or Context Omission: Vague or underspecified input leading to multiple plausible interpretations. 

Figure~\ref{fig:error-types} shows that Zero-Shot Prompt and NL2TL suffer most from temporal misinterpretation and loss of nesting, each accounting for nearly half of their total errors. This demonstrates the limitations of structure-unaware methods in handling deep logic.
NL2SPEC offers modest improvements but still exhibits instability across categories. In contrast, \textsc{Req2LTL} effectively eliminates errors in the first four categories. Its only remaining issues are due to inherent ambiguity in the input, which is difficult to resolve without contextual grounding.

In summary, compared with existing prompting-based methods, 
\textsc{Req2LTL} demonstrates clear advantages on classes of requirements 
that are particularly challenging: 
(1) implicit temporal semantics (e.g., ``will be set'', ``unless''), 
(2) deeply nested logic structures (63.2\% of industrial dataset), 
(3) pattern-triggered constraints that require correct scope and causality, 
and (4) long clause dependencies that often exceed the syntactic modeling 
capacity of end-to-end LLMs. 
By enforcing structured semantic decomposition, \textsc{Req2LTL} 
progressively translates complex sentences and effectively avoids these errors.

\smallskip
\noindent\textbf{The answer to RQ4:} 
Structural challenges are prevalent in real-world industrial requirements, and \textsc{Req2LTL} effectively mitigates them by enforcing semantic structure and reducing error propagation during generation.
Nonetheless, certain cases of linguistic ambiguity persist, motivating the integration of visualized feedback and human-in-the-loop correction, which we explore in the next section.

\subsection{Threats to Validity}
One potential threat is data leakage in publicly available academic benchmarks, which may have been seen during pretraining by LLMs. However, our industrial dataset, composed of proprietary aerospace requirements, was never part of any model's training corpus, ensuring unbiased evaluation.
Another limitation lies in domain scope: while results are strong in general scenarios and the aerospace field, generalization to other domains is untested.
Finally, the quality of all baselines and \textsc{Req2LTL} depends on the behavior of the underlying LLM, which may vary across model versions or providers.
In our experiment, we conducted a fairness treatment.



\section{Case Study and Discuss}
\subsection{Representative Failure Cases}
\label{sec:9.1}

While \textsc{Req2LTL} performs well overall, certain industrial requirements continue to pose challenges.
The most common failure type we observe is \textit{ambiguity or context omission}, where the intended timing of actions is not explicitly stated, making precise formalization unreliable.

\textbf{Example, Req04}:
\textit{``The control system should as soon as possible initiate the heading adjustment function upon receiving a verified ARINC 429 waypoint command, ultimately reducing the deviation angle to less than 2 degrees.''}

This requirement specifies two subgoals: triggering the heading adjustment promptly and eventually reducing the deviation. However, the phrase ``\textit{as soon as possible}'' is vague. The model assigns both subgoals the $F$ (eventually) operator:

{
\small
\[
\begin{aligned}
\text{\textcolor{warmdarkred}{LTL04}: } G(&\,\text{WaypointCmd} = \text{True} \rightarrow \\
  &\, (\textcolor{warmdarkred}{F}\,(\text{HeadingFun} = \text{True}) \land F\,(\text{DevAngleLow} < 2)))
\end{aligned}
\]
}

The first sub-goal should instead use $X$, indicating that the action is to be executed at the next time point.
The misclassification stems from the natural language's lack of explicit timing, which the model cannot resolve without domain-specific understanding.

\smallskip
\noindent \textbf{Discussion:}
This case highlights a common challenge in industrial requirement translation: natural language descriptions may contain inherent ambiguity or omit critical information. For instance, phrases like \textit{as soon as possible} suggest urgency but lack formal definition. Without clear temporal anchors or domain-specific context, even structure-aware models tend to default to loose eventualities such as $F$, resulting in misaligned formalizations.
More broadly, this reflects a limitation in the requirement itself. Many industrial specifications rely on shared engineering knowledge and are written with implicit assumptions, rather than precise, machine-interpretable semantics. In such cases, automated systems struggle to infer intent or resolve underspecified constructs.

This inherent ambiguity cannot be fully resolved by automated translation alone, even when guided by structured prompts or intermediate representations. This underscores the need for human-in-the-loop mechanisms. \textsc{Req2LTL} addresses this by exposing editable intermediate structures through \textit{OnionL}, allowing engineers to inspect, refine, and correct semantic errors. In the next section, we show how this feedback loop enables low-effort repair of misclassifications while retaining automation efficiency.


\subsection{Structured Feedback via Visualized \textit{OnionL}}


To facilitate error correction in cases of semantic ambiguity, \textsc{Req2LTL} provides a visualized \textit{OnionL} representation to support user inspection and guided refinement. As illustrated in Figure~\ref{fig:onionl-visual}, the model’s intermediate output is rendered as a structured diagram, enabling engineers to directly examine how a requirement has been semantically interpreted. When inconsistencies are identified, users may either revise the temporal operators within the diagram or provide natural-language feedback to regenerate a corrected structure. 
In the example Req04 and LTL04, replacing the first $F$ with $X$ in the tree was sufficient to recover the correct semantics, producing the following LTL formula. This correction process is both lightweight and non-intrusive, requiring no modifications to the original input and preserving the automation pipeline while allowing precise semantic control.

{
\small
\[
\begin{aligned}
\text{\textcolor{darkgreen}{LTL04}: } G(&\,\text{WaypointCmd} = \text{True} \rightarrow \\
  &\, (\textcolor{darkgreen}{X}\,(\text{HeadingFun} = \text{True}) \land F\,(\text{DevAngleLow} < 2)))
\end{aligned}
\]
}

\smallskip
\noindent\textbf{Discussion:}
The visualized \textit{OnionL} structure is not merely an inspection aid—it serves as a pivotal component for enabling human-in-the-loop correction. Users can directly manipulate the tree or interact with the underlying LLM to generate revised outputs, forming a feedback loop that enhances both robustness and interpretability. This mechanism ensures traceability, supports auditability, and aligns with certification workflows, making \textsc{Req2LTL} a practical tool for safety-critical engineering contexts.
Its effectiveness was confirmed during expert evaluation: among 112 industrial requirements, 13 LTL formulas initially exhibited semantic errors. All were corrected within 10 minutes using the visual interface, demonstrating the practicality of the approach in real-world formalization scenarios.
It should be emphasized that these 13 manually corrected cases were used solely to demonstrate the practicality of the visualization-based human-in-the-loop interface. All quantitative results reported in RQ1 and RQ2 were obtained from the fully automated pipeline with manual validation disabled, ensuring that the performance metrics strictly reflect automated capability.


\section{Related Works}
Early approaches~\cite{swick2024flexible, Tellex2011ApproachingTS, xu2024ltl, buzhinsky2019formalization, wu2024survey, Zhang2020AutomatedGO} for translating natural language (NL) requirements into formal specification relied heavily on handcrafted rules, such as syntactic preprocessing, pattern matching, and attribute grammars. While effective in constrained domains, these methods lack scalability and robustness due to their domain-specific assumptions and limited expressiveness.
The advent of neural models introduced data-driven paradigms. Seq2Seq architectures~\cite{Gopalan2018SequencetoSequenceLG_ref17}, semantic parsers~\cite{Wang2020LearningAN, He2021DeepSTLF}, and template-guided generators enabled automatic learning from paired NL–LTL datasets. However, these models often generalize poorly to real-world requirements that exhibit implicit semantics, nested logic, and domain-specific terminology, especially when training data lacks such complexity.

Transformer-based language models~\cite{Vaswani2017AttentionIA} such as GPT~\cite{Madotto2020LanguageMAGPT}, T5~\cite{Raffel2019ExploringTLT5}, and PaLM~\cite{Chowdhery2022PaLMSL} brought significant improvements in text generation~\cite{Chintagunta2021MedicallyAG} and code synthesis~\cite{Chen2021EvaluatingLLPython_ref6}. Building on these capabilities, recent methods~\cite{Chen2023NL2TLTN, Cosler2023nl2specIT, Fuggitti2023NL2LTLA} have explored few-shot prompting, abstraction templates, and fine-tuning for formal specification generation. More recent work~\cite{wang2025conformalnl2ltl, zhao2024nl2ctl, mendoza2024synttl} incorporates interactive feedback and decomposition mechanisms to enhance coverage and interpretability across a broader range of scenarios.
Nonetheless, LLM-based methods remain sensitive to prompt phrasing and often lack structural transparency. They struggle with faithfully modeling compositional semantics and handling ambiguity in complex industrial requirements. These challenges point to the need for more robust, structure-aware translation frameworks capable of preserving semantic fidelity under domain-specific constraints.

\section{Conclusion}
This paper introduced \textsc{Req2LTL}, a modular framework for translating natural language requirements into LTL specifications. 
The framework introduces a novel intermediate representation \textit{OnionL}, which decomposes requirements into a compositional tree composed of semantic scopes, logical relations, and atomic propositions, thereby enabling structured and verifiable translation. 
By combining LLMs for hierarchical semantic decomposition with a deterministic, rule-based translator for validation and LTL conversion, \textsc{Req2LTL} achieves 88.4\% semantic accuracy and 100\% syntactic correctness on real-world aerospace requirements, substantially outperforming prior approaches. 
Our results demonstrate that \textsc{Req2LTL} effectively bridges the gap between informal requirements and formal specifications, and provides a practical foundation for scalable adoption in safety-critical industrial systems.
In future work, we plan to extend the framework to more expressive temporal logics such as STL and MTL.

\section{Data Availability}

Due to confidentiality restrictions related to aerospace data and the ongoing integration of the tool into our industrial partner's internal verification platform, the full source code and industrial dataset cannot be released at this stage. With partner’s approval, we plan to release a standalone prototype of \textsc{Req2LTL}, together with portions of the aerospace dataset and an interactive front-end interface, once the integration work is complete. 
In the meantime, a demonstration package of \textsc{Req2LTL}, including a usage guide and video, is publicly available at 
\url{https://github.com/Meng-Nan-MZ/Req2LTL.git}.


\section*{Acknowledgements}
This work was supported in part by the National Natural Science Foundation of China (Nos. 62192730, 62192734, 62302375, 62472339, 6240073166), the China Postdoctoral Science Foundation funded project (No. 2023M723736), the Outstanding Youth Science Foundation of Shaanxi Province under Grant 2025JC-JCQN-083, the Fundamental Research Funds for the Central Universities (No. XJSJ24086), and the CCF-Huawei Populus Grove Fund (No. CCF-HuaweiFM202507).



\clearpage

\bibliographystyle{IEEEtran}
\bibliography{citation}

\end{document}